\documentclass[letter]{aa}

\usepackage{graphicx}
\usepackage{txfonts}
\usepackage{hyperref}
\usepackage{xcolor}
\hypersetup{
colorlinks,
linkcolor={red!80!black},
citecolor={blue!80!black},
urlcolor={blue!80!black}
}

\newenvironment{myfont}{\fontfamily{ppl}\selectfont}{\par}
\DeclareTextFontCommand{\textmyfont}{\myfont}

\newcommand{\code}[1]{\texttt{#1}}

\def\nifs{\iso{56}Ni}

\def\cm3{cm$^{-3}$}
\def\kms{\mbox{km~s$^{-1}$}}

\def\msun{$M_{\odot}$}

\def\one{\ts {\,\sc i}}
\def\two{\ts {\,\sc ii}}
\def\three{\ts {\,\sc iii}}

\def\beq{\begin{equation}}
\def\eeq{\end{equation}}

\def\lesssim{\mathrel{\hbox{\rlap{\hbox{\lower4pt\hbox{$\sim$}}}\hbox{$<$}}}}
\def\gtrsim{\mathrel{\hbox{\rlap{\hbox{\lower4pt\hbox{$\sim$}}}\hbox{$>$}}}}

\def\one{{\,\sc i}}
\def\two{{\,\sc ii}}
\def\three{{\,\sc iii}}

\def\v1d{{\code{V1D}}}
\def\sumo{{\code{SUMO}}}
\def\kepler{{\code{KEPLER}}}

\def\cmfgen{{\code{CMFGEN}}}

\def\oidoub{[O\one]\,$\lambda\lambda$\,$6300,\,6364$}
\def\caiidoub{[Ca\two]\,$\lambda\lambda$\,$7291,\,7323$}

\newcommand{\iso}[2]{\ensuremath{^{#1}\rm{#2}}}

\begin{document}

   \title{Radiative-transfer modeling of supernovae in the nebular-phase. \\
A novel treatment of chemical mixing in spherical symmetry.}

   \titlerunning{SN nebular-phase modeling}

\author{Luc Dessart\inst{\ref{inst1}}
  \and
   D. John Hillier\inst{\ref{inst2}}
  }

\institute{
Institut d'Astrophysique de Paris, CNRS-Sorbonne Universit\'e, 98 bis boulevard Arago, F-75014 Paris, France.\label{inst1}
\and
    Department of Physics and Astronomy \& Pittsburgh Particle Physics,
    Astrophysics, and Cosmology Center (PITT PACC),  \hfill \\ University of Pittsburgh,
    3941 O'Hara Street, Pittsburgh, PA 15260, USA.\label{inst2}
  }

   \date{}

  \abstract{
Supernova (SN) explosions, through the metals they release, play a pivotal role in the chemical evolution of the Universe and the origin of life. Nebular phase spectroscopy constrains such metal yields, for example through forbidden line emission associated with O\one, Ca\two, Fe\two, or Fe\three. Fluid instabilities during the explosion produce a complex 3D ejecta structure, with considerable macroscopic, but no microscopic, mixing of elements. This structure sets a formidable challenge for detailed nonlocal thermodynamic equilibrium radiative transfer modeling, which is generally limited to 1D in grid-based codes.  Here, we present a novel and simple method that allows for macroscopic mixing without any microscopic mixing, thereby capturing the essence of mixing in SN explosions. With this new technique, the macroscopically mixed ejecta is built by shuffling in mass space, or equivalently in velocity space, the shells from the unmixed coasting ejecta. The method requires no change to the radiative transfer, but necessitates high spatial resolution to resolve the rapid variation in composition with depth inherent to this shuffled-shell structure. We show results for a few radiative-transfer simulations for a Type II SN explosion from a 15\,\msun\ progenitor star. Our simulations capture the strong variations in temperature or ionization between the various shells that are rich in H, He, O, or Si. Because of nonlocal energy deposition, $\gamma$ rays permeate through an extended region of the ejecta, making the details of the shell arrangement unimportant. The greater physical consistency of the method delivers spectral properties at nebular times that are more reliable, in particular in terms of individual emission line strengths, which may serve to constrain the SN yields and, for core collapse SNe, the progenitor mass. The method works for all SN types.
}

   \maketitle

\section{Introduction}

The complicated 3D structure observed in supernova (SN) remnants from Cas A \citep{fesen_casA_06} to the young SN\,1987A \citep{abellan_87A_17}, or the detection of polarized radiation from Type II SNe \citep{shapiro_sutherland_82,leonard_04dj_06} provide evidence that even standard core-collapse SNe are inherently asymmetric and heterogeneous.  The complex 3D distribution of elements likely arises from the neutrino-driven explosion mechanism combined with fluid instabilities triggered by the propagation of the SN shock across the progenitor envelope \citep{fryxell_mueller_arnett_91,kifonidis_00,wongwathanarat_15_3d,ono_3d_20}. This structure contrasts with the progenitor chemical distribution, which is thought to exhibit stacked shells of distinct composition (this picture is being challenged by recent simulations; \citealt{couch_3d_presn_15}), and with an increasing mean atomic weight towards the denser inner layers of the star, culminating with the Fe core in the innermost layers when the massive star is ripe for core collapse (see, e.g., \citealt{arnett_book_96}).

 \begin{figure*}
\centering
\includegraphics[width=0.495\hsize]{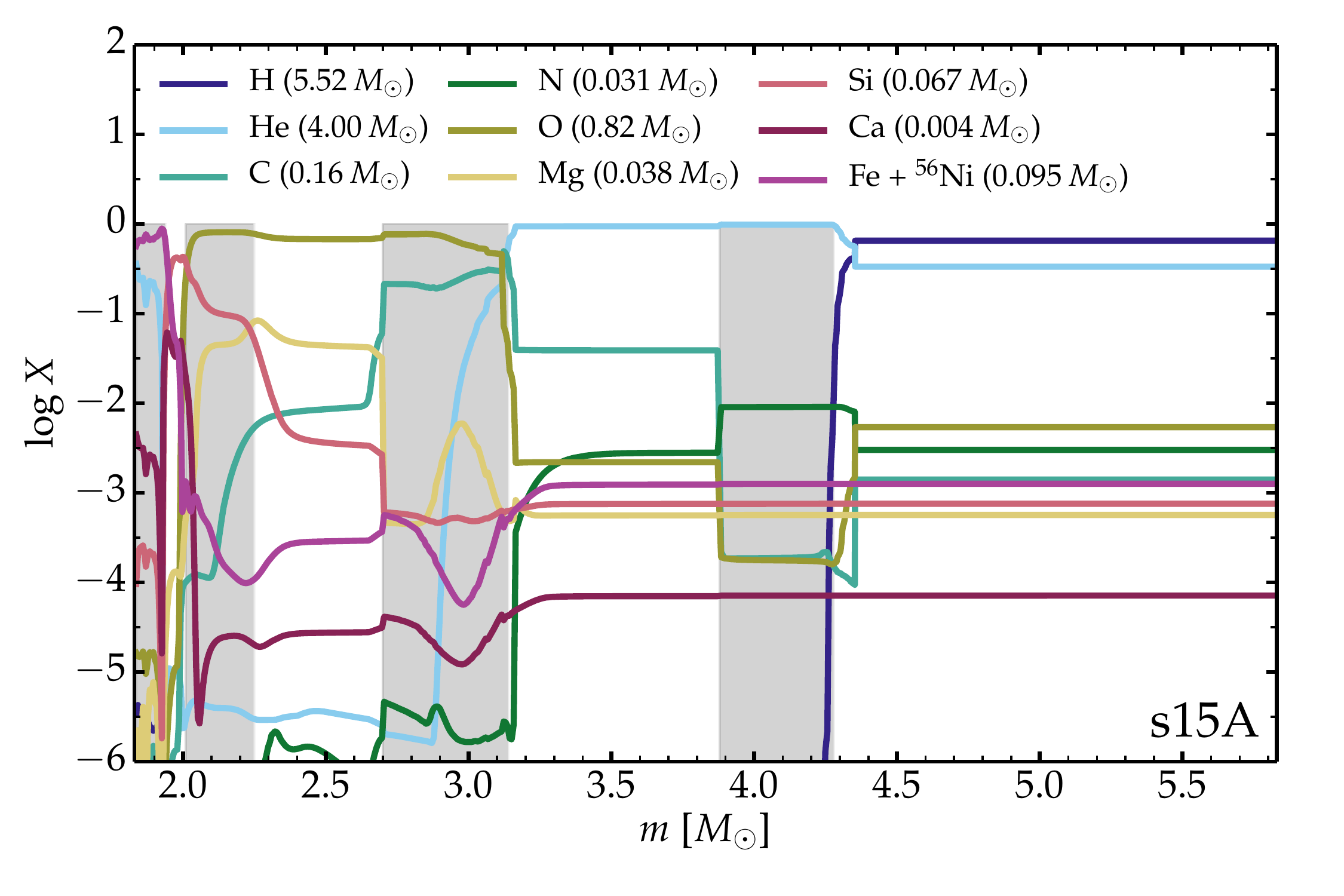}
\includegraphics[width=0.495\hsize]{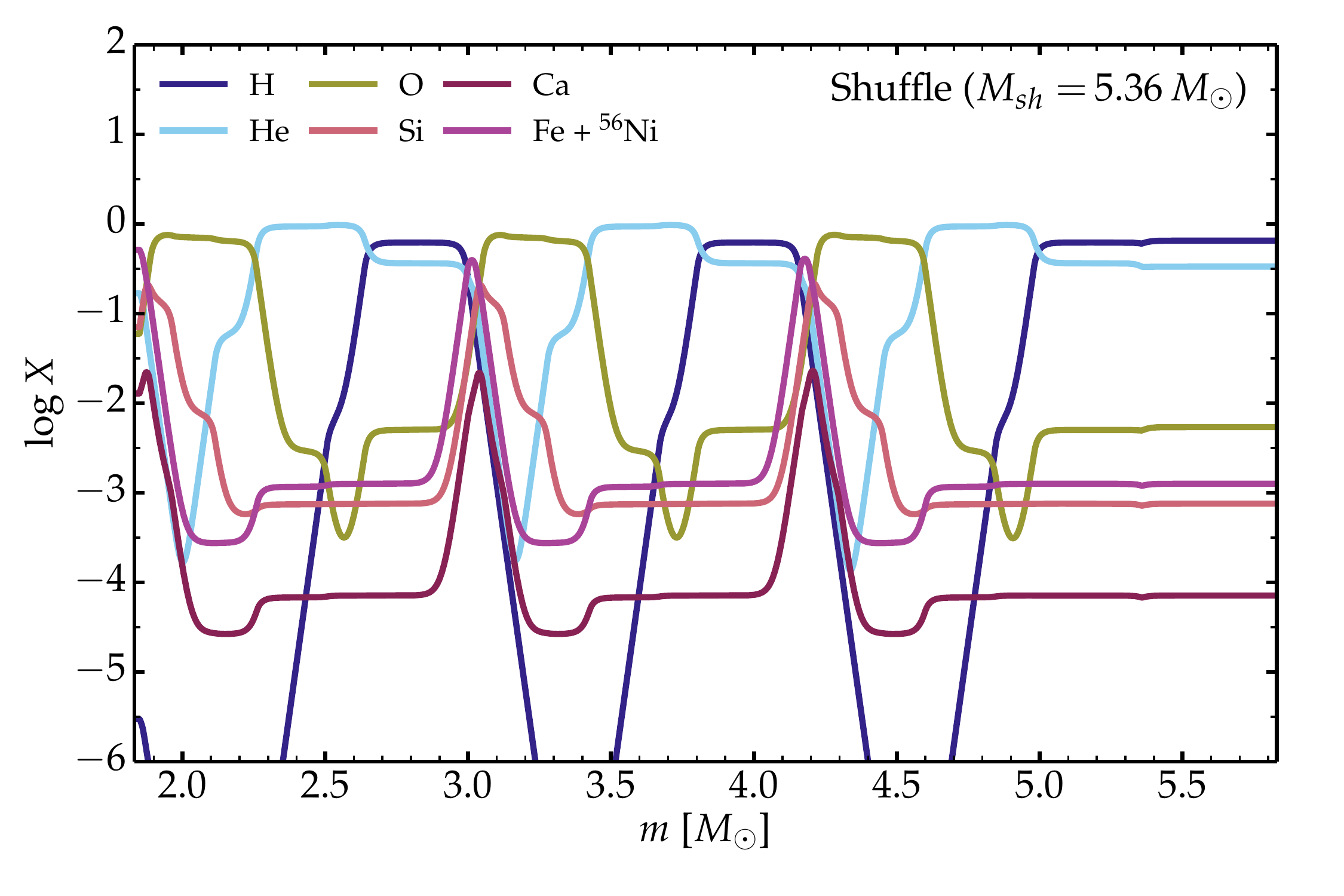}
\vspace{-0.4cm}
\caption{{\it Left:} Unmixed chemical composition versus Lagrangian mass in the s15A ejecta model of \citet{WH07} at 345\,d (the original mass fraction is shown for \nifs), arising from a 15\,\msun\ progenitor star on the main sequence, and having an ejecta kinetic energy of $1.2 \times 10^{51}$\,erg, a total ejecta mass of 10.96\,\msun, and 0.083\,\msun\ of \nifs\ initially. To properly identify the eight shells, we also show the profiles for C, N, Mg, and Ca. {\it Right:} Composition profile obtained with the new mixing approach presented in this letter, using  $M_{\rm sh}=$\,5.36\,\msun, in which only macroscopic mixing is applied by splitting and shuffling ejecta shells identified in the unmixed model shown at left. In both panels (and in Fig.~\ref{fig_sn_comp}), the origin of the $x$-axis is at 1.83\,\msun, which corresponds to the mass of the compact remnant.
\label{fig_prog_comp}
}
\end{figure*}

 The complex 3D ejecta structure sets a challenge for the radiative transfer modeling of SN radiation. The common expedient assumes an overall spherical symmetry of the homologously expanding ejecta, but with an enhanced mixing of all species. This mixing is both macroscopic (material from low velocity is advected out to large velocity, and vice versa) and microscopic (the ``advected" material is fully mixed with the material at its new location). We refer to this approach as the ``old", standard, mixing technique (see, e.g., \citealt{D15_SNIbc_I}). This approximation is adequate for bolometric light-curve calculations. However, because the composition is altered in a nonphysical manner, the opacity and emissivity of the plasma is no longer accurate, which may impact the SN color, the multiband light curves, and the spectral properties.

 The shortcomings of a combined macroscopic and microscopic mixing are best seen in nebular-phase spectra. At late times, the plasma cools primarily through line emission, and their cooling power depends sensitively on the local plasma composition. As is well known \citep{fransson_chevalier_89}, and recently demonstrated with detailed calculations \citep{DH20_neb}, mixing Ca with O-rich material affects \oidoub\ emission.  The sensitivity of line strengths to microscopic mixing, vividly illustrated by O\one, also holds for other emission lines. While some mixing may occur during the ultimate stages of massive star evolution (see, e.g., \citealt{collins_presn_18}), it should not be enforced artificially by the need for a simplistic treatment of mixing driven by numerical convenience. Instead, reliable inferences from nebular-phase spectra require an accurate description of chemical mixing in SN ejecta.

\section{Treatment of chemical mixing: standard and new approaches}

To avoid mixing material from different shells of the progenitor star, one could treat each dominant pre-SN shell (i.e., the O/Ne/Mg shell, or the He/N shell etc) independently of the other shells, essentially one at a time. Each shell may be spread in velocity space and given a prescribed decay power. With a proper bookkeeping, the method can be set to preserve mass, energy, match the total decay power, and provide a complete nebular-spectrum for a given SN ejecta. But by treating each dominant shell independently of the others, any cross-talk between shells is neglected. We have tried various approaches along this route for Type II-Plateau SNe but all failed to deliver a reasonably looking nebular spectrum. The reason is that, even at 200\,$-$\,300\,d after explosion, there is considerable reprocessing by the whole ejecta of photons injected below 5000\,\AA\ (e.g., photons emitted from the hotter He-rich shell material; see also \citealt{jerkstrand_15_iib}). The spectrum formation is a global process which requires the modeling of the full ejecta and associated couplings.

\begin{figure}
\centering
\includegraphics[width=\hsize]{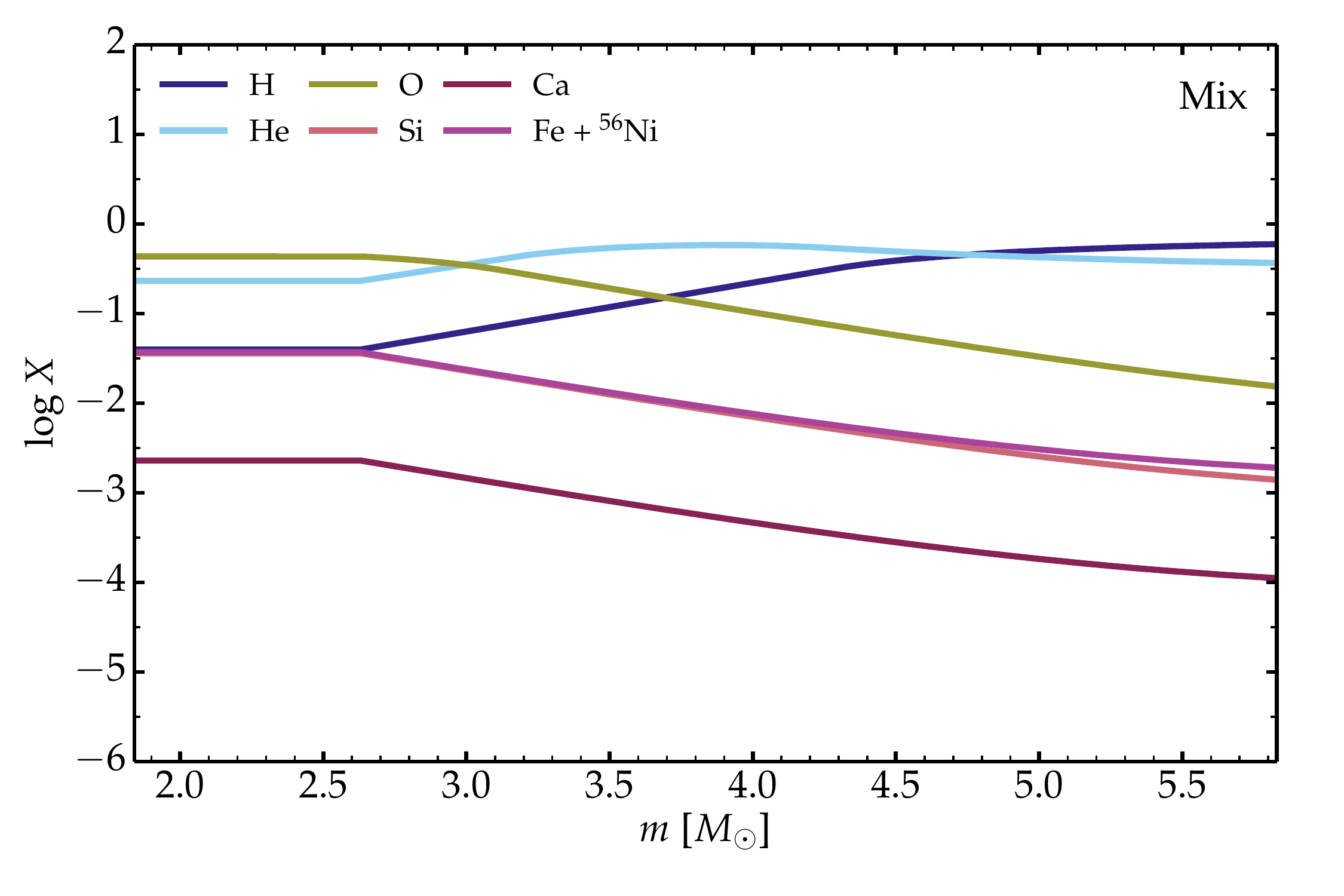}
\vspace{-0.7cm}
\caption{Counterpart to the right panel of Fig.~\ref{fig_prog_comp}, but now using the old, standard, approach for mixing in which the chemical mixing is microscopic and macroscopic and applied to all species  (see, e.g., \citealt{D15_SNIbc_I}). The complex composition profiles of Fig.~\ref{fig_prog_comp} are completely lost with this simplistic and indiscriminate mixing technique.
\label{fig_sn_comp}
}
\end{figure}

In this letter, we present a novel treatment of chemical mixing in spherical symmetry, applicable to all SN types, and which requires no adjustment to the radiative transfer,  here performed with the nonlocal thermodynamic equilibrium (nonLTE) radiative transfer  code \cmfgen\ \citep{HD12}. The adjustment applies instead to the initial conditions for the coasting ejecta. Since the SN ejecta structure cannot be broken laterally in 1D (by having material of distinct composition coexist at the same velocity), the fundamental idea of the method is to break this structure radially.  Taking the unmixed SN ejecta, we select the region that covers from the innermost ejecta layers rich in metals to somewhere beyond the base of the H-rich shell, and bounded by the Lagrangian mass $M_{\rm sh}$. We then split each of the eight dominant shells (concisely tagged H/He, He/N, He/C, O/C, O/Ne/Mg, O/Si, Si/Ca, and Fe/He) into three equal parts and shuffle these 24 sub-shells within the same region limited by $M_{\rm sh}$. Because the ejecta is coasting, this shuffling in mass space is equivalent to a shuffling in radial or velocity space. In this brute-force approach, the only requirement for \cmfgen\ is to increase the number of grid points in order to resolve the rapid variation of composition with depth. Rather than using a total of $\sim$\,80 grid points for a standard nebular-phase model (as in the simulations presented in \citealt{DH20_neb}), we require $\sim$\,400 grid points, of which 90\% are devoted to resolving the ejecta layers within $M_{\rm sh}$. A detailed presentation of the procedure is given in the Appendix~A.

\begin{figure*}
\centering
\includegraphics[width=\hsize]{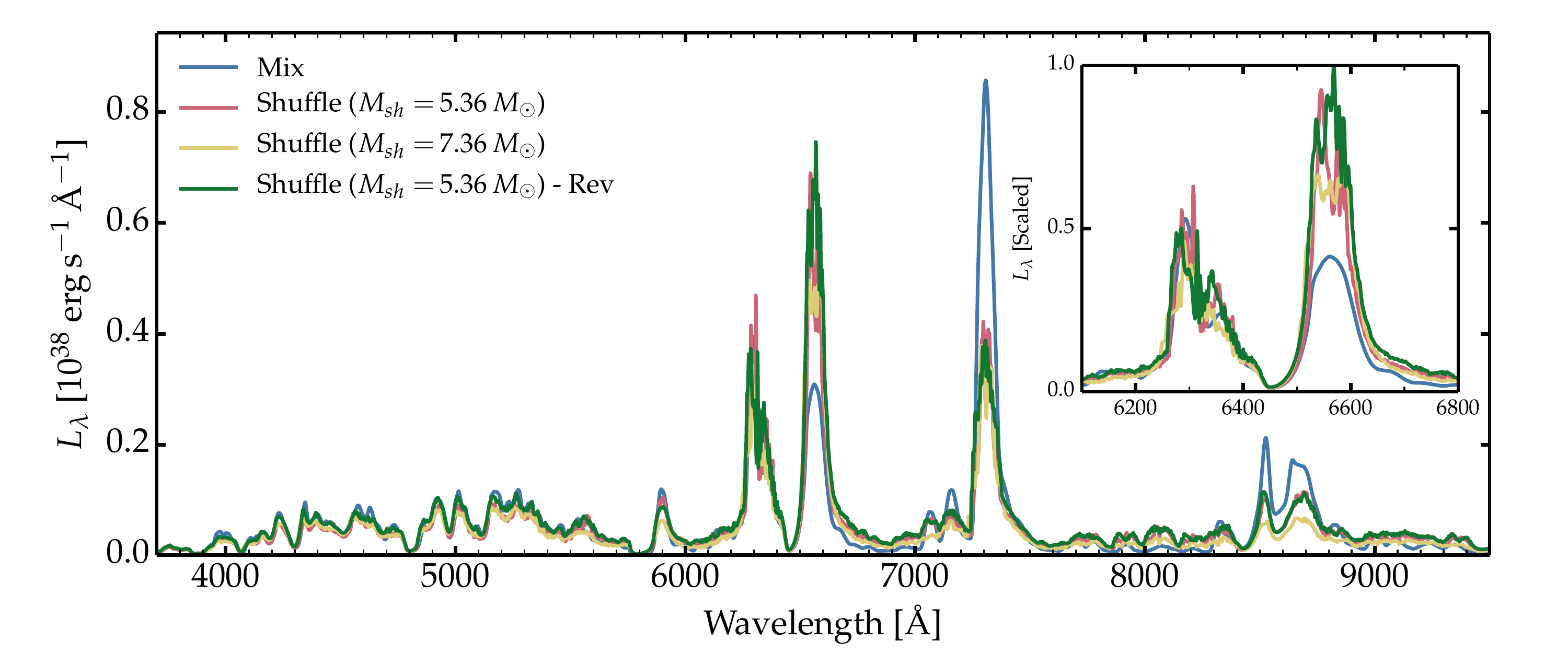}
\vspace{-0.6cm}
\caption{Comparison of the optical luminosity (no normalization is applied) for models using the standard (``Mix") and the new mixing technique based on the shuffling of shells (``Shuffle", with various choices for $M_{\rm sh}$; the fine, resolved, structure in the spectra is real and caused by the rapid variation in composition in these shuffled-shell models). For model with suffix ``Rev", the shell stacking order is reversed. While the different models with shuffling appear similar (highlighting some degeneracy with mixing), there is a strong difference with the results obtained with the standard mixing technique, in particular for H$\alpha$ and \caiidoub.
}
\label{fig_spec}
\end{figure*}

To illustrate the procedure, Fig.~\ref{fig_prog_comp} shows the undecayed composition versus Lagrangian mass  in the s15A ejecta model at 345\,d computed with \kepler\ \citep{WH07}. The eight main shells are indicated with an alternate color of grey and white: starting at the remnant mass of 1.83\,\msun, one goes through the shell rich in Fe/He, followed by the Si/Ca shell, the O-rich shell (which we split into three distinct shells, namely O/Si, O/Ne/Mg, and O/C), the He-rich shell (which we split into two distinct shells, namely He/C and He/N), and finally the H/He shell from the progenitor H-rich envelope (the thin He/N/H shell is treated here as being part of the massive H/He shell) all the way to 12.79\msun, which is the model total mass at collapse.

Splitting each of the eight shells (of distinct composition; the outermost shell involved in the process is limited to the range 4.28 to 5.36\,\msun, while beyond 5.36\,\msun, the H-rich material is left untouched) in three equal-mass sub-shells, we cycle three times through these eight shells and stack them on top of each other, starting from the ejecta base. The result is similar to what is shown in the left panel of Fig.~\ref{fig_prog_comp} but now the pattern of eight shells is repeated three times with all eight shells being one third of their original mass. This shuffling thus preserves the ejecta mass and kinetic energy but introduces the macroscopic mixing we need. There is no microscopic mixing between shells.

The ejecta composition produced with the new mixing technique is shown for the main species (i.e., H, He, O, Si, and Fe) in the right panel of Fig.~\ref{fig_prog_comp}. For comparison, we show in Fig.~\ref{fig_sn_comp} the results obtained with the standard mixing technique: the detailed composition segregation is lost in that case, all species being present throughout the ejecta, exhibiting a broad and smooth profile. The new approach avoids these pitfalls and retains the numerical advantages of spherical symmetry. In addition, the \nifs\ is no longer mixed throughout the ejecta so \nifs-rich material has a greater density, causing a greater Fe absorption (this property may be mitigated by the \nifs\ bubble effect, which is the expansion caused by decay heating; e.g., \citealt{woosley_87A_late_88}).

The new mixing technique is flexible since one can design many different shell arrangements to mimic macroscopic mixing. Rather than stacking the sub-shells from the ejecta base, one may start from $M_{\rm sh}$ and progress inwards towards the ejecta base. This choice puts some \nifs\ material at larger velocity (the innermost \nifs\ shell would be located at 5.36\,\msun\ in Fig.~\ref{fig_sn_comp}, while H-rich material would then be present at the lowest ejecta velocities). We may also increase $M_{\rm sh}$ to allow a greater mixing of the H-rich material with the metal-rich regions. Shells could also be split into more than three parts but thinner shells are harder to resolve numerically. On the other hand, splitting the shells in two parts may produce too little mixing. Finally, we could also mix some \nifs-rich material to large velocities, beyond $M_{\rm sh}$, by merely switching a fraction of the \nifs-rich shell in the inner ejecta with an H-rich shell of the same mass in the outer ejecta. Results for some of these configurations are presented below.

While the adopted mixing in Fig.~\ref{fig_sn_comp} may seem weak, the shuffled-shell structure covers the $4\pi$ angle, mimicking the presence of numerous ``clumps" of the same composition at the same velocity (or mass coordinate). Additional shifts along different directions, which cannot occur in 1D,  are not critical since the $\gamma$-rays from radioactive decay fill rather uniformly the entire mixed region -- their typical mean free path of $\sim 5 \times 10^{15}$\,cm in model s15A at 345\,d is comparable to the SN radius.   Hence, allowing for more shells or even doing a 3D treatment would merely randomize the location of emission and absorption of the ejecta material, but this would merely produce smoother line profiles while not changing much the physics at work. The impact on total line fluxes and such is expected to be small. Although the mixing approach is distinct from that used in the Monte Carlo code \sumo\ \citep{jerkstrand_87a_11}, the resulting 3D structure and its influence on the radiative transfer are similar.

Using the coasting ejecta model s15A described above (see also \citealt{WH07}), we implement various types of chemical mixing and investigate the resulting radiative properties with \cmfgen.  The \cmfgen\ calculations assume steady state, a SN age of 345\,d, and treat the ejecta composition in detail (Fig.~\ref{fig_sn_comp}). We include up to three ionization stages for H, He, C, N, O, Ne, Na, Mg, Al, Sc, Si, S, Ar, K, Ca, Ti, Cr, Fe, Co, and Ni, and only the \nifs-decay chain is considered.  The ejecta inside $M_{\rm sh}$, bounded by velocities between 250 and 2500\,\kms, is resolved uniformly with 320 points, thus with a fixed resolution of 7\,\kms. Between 2500\,\kms\ and the maximum velocity of 7000\,\kms, we employ 30 points equally spaced on an optical-depth scale.

\section{Results}

Figure~\ref{fig_spec} compares the optical spectra (shown as a luminosity $L_\lambda$, and thus without any scaling) for the \cmfgen\ calculations based on the standard technique for mixing (i.e., which is both microscopic and macroscopic), and various incarnations of the new mixing technique that shuffles ejecta shells in mass (or in velocity) space. Although all four models have the same \nifs\ mass, the different mixing introduces variations in decay-power absorbed (and thus bolometric luminosity) at the 10\% level.

\begin{figure}
\centering
\includegraphics[width=\hsize]{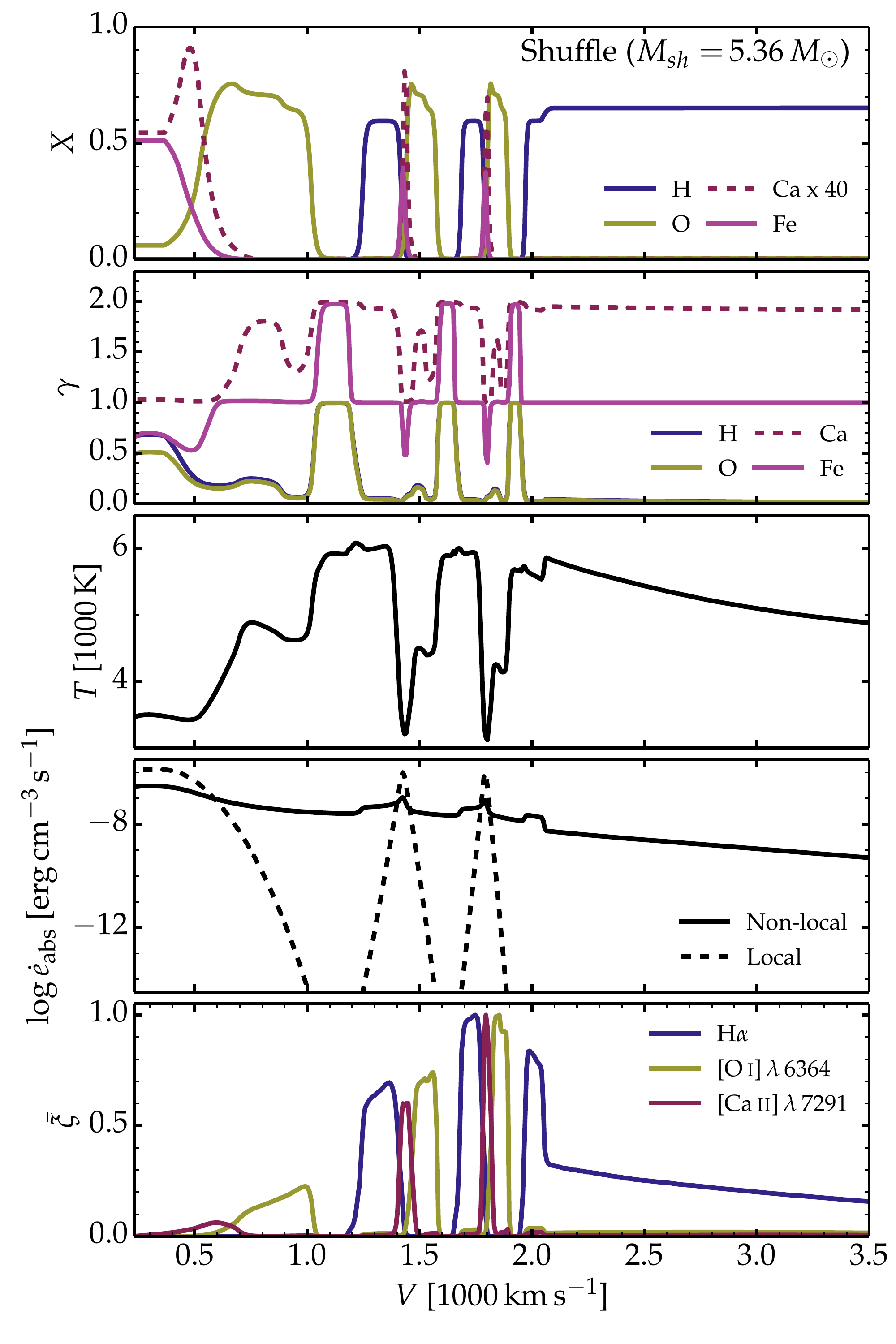}
\vspace{-0.7cm}
\caption{Illustration of ejecta and radiation properties for the shuffled-shell model with $M_{\rm sh}=5.36$\,\msun. From top to bottom, we show the profiles versus velocity for the mass fraction of H, O, Ca, and Fe, their ionization state (zero for neutral, one for once ionized etc), the gas temperature, the decay power absorbed for local (shown for comparison only) and nonlocal deposition, and the formation regions for H$\alpha$, [O\,{\sc i}]\,$\lambda$\,6364, and [Ca\,{\sc ii}]\,$\lambda$\,7291 (the quantity  $\int \xi \, d\log V $ is the line equivalent width, which is here normalized to unity for visibility).
}
\label{fig_shuffle_ej_prop}
\end{figure}

Two striking results emerge from Fig.~\ref{fig_spec}, whose interpretation can be facilitated by inspection of the ejecta and radiation properties for one of the shuffled-shell models shown in Fig.~\ref{fig_shuffle_ej_prop}. First, the main difference between models is limited to H$\alpha$ and \caiidoub, with little impact on \oidoub. It is complicated to identify the core reason for this because the physics is very nonlinear and complex. With the standard mixing technique, the decay power is absorbed by a nearly homogeneous metal-rich inner ejecta, from which [Ca\two] forbidden lines perform most of the cooling. In contrast, in the new shuffled-shell approach, the [Ca\two] forbidden lines form mostly within the Si-rich layer (Fig.~\ref{fig_shuffle_ej_prop}). Being of low mass (see Fig.~\ref{fig_prog_comp}), this Si-rich layer absorbs a small fraction of the total decay power, which drastically limits the maximum flux in [Ca\two]\,$\lambda\lambda$\,$7291,\,7323$. In the ``Mix'' model, this maximum flux can be huge, and potentially equal to the total decay power available, because the Ca abundance is  allowed to be large throughout the ejecta (Fig.~\ref{fig_sn_comp}).

The second striking result is that the various choices for the shuffling of shells do not lead to much diversity. The spectral properties are therefore weakly sensitive to the details of mixing (as long as the mixing of \nifs\ is comparable; see \citealt{DH20_neb}), which is poorly constrained in SN ejecta. This degeneracy arises because the emergent radiation is primarily dependent on the decay power absorbed by each shell. By allowing for the presence of \nifs-rich shells  at various locations out to about 2000\,\kms, nonlocal energy deposition allows decay power to permeate nearly uniformly throughout this extended region (see Fig.~\ref{fig_shuffle_ej_prop}) -- the specific location of the various other shells becomes irrelevant. Furthermore, the key physics operating in each of these shells (opacity, emissivity) is primarily dependent on the shell mass and composition, which is the same in all three incarnations shown in Fig.~\ref{fig_spec} since it is fixed for a given progenitor (here the unmixed s15A model).

Although one may argue that the standard mixing technique and our new technique yield similar-looking spectra, they are quantitatively different, and the new technique has physical consistency that the standard technique was severely lacking. With the new technique, we can rely on our results, which are no longer biased by an unphysical mixing of species.  For example, our new simulations yield very complex temperature and ionization profiles, reflecting the different composition and coolants in the various shells. We also have confidence on the origin of the emission from forbidden lines, while the treatment of the full ejecta takes full consideration of any cross-talk between individual shells. In the shuffled-shell model with $M_{\rm sh}=$\,5.36\,\msun, the H$\alpha$ line forms throughout the H-rich ejecta layers (down to the innermost layer which is placed around 1400\,\kms), the \oidoub\ line forms primarily in the O-rich shell, while  \caiidoub\ forms in the Si-rich shell, where Ca is neutral -- over-ionization of Ca is in part responsible for the lack of Ca\two\ emission from the outer H-rich ejecta layers.

\begin{figure}
\centering
\includegraphics[width=\hsize]{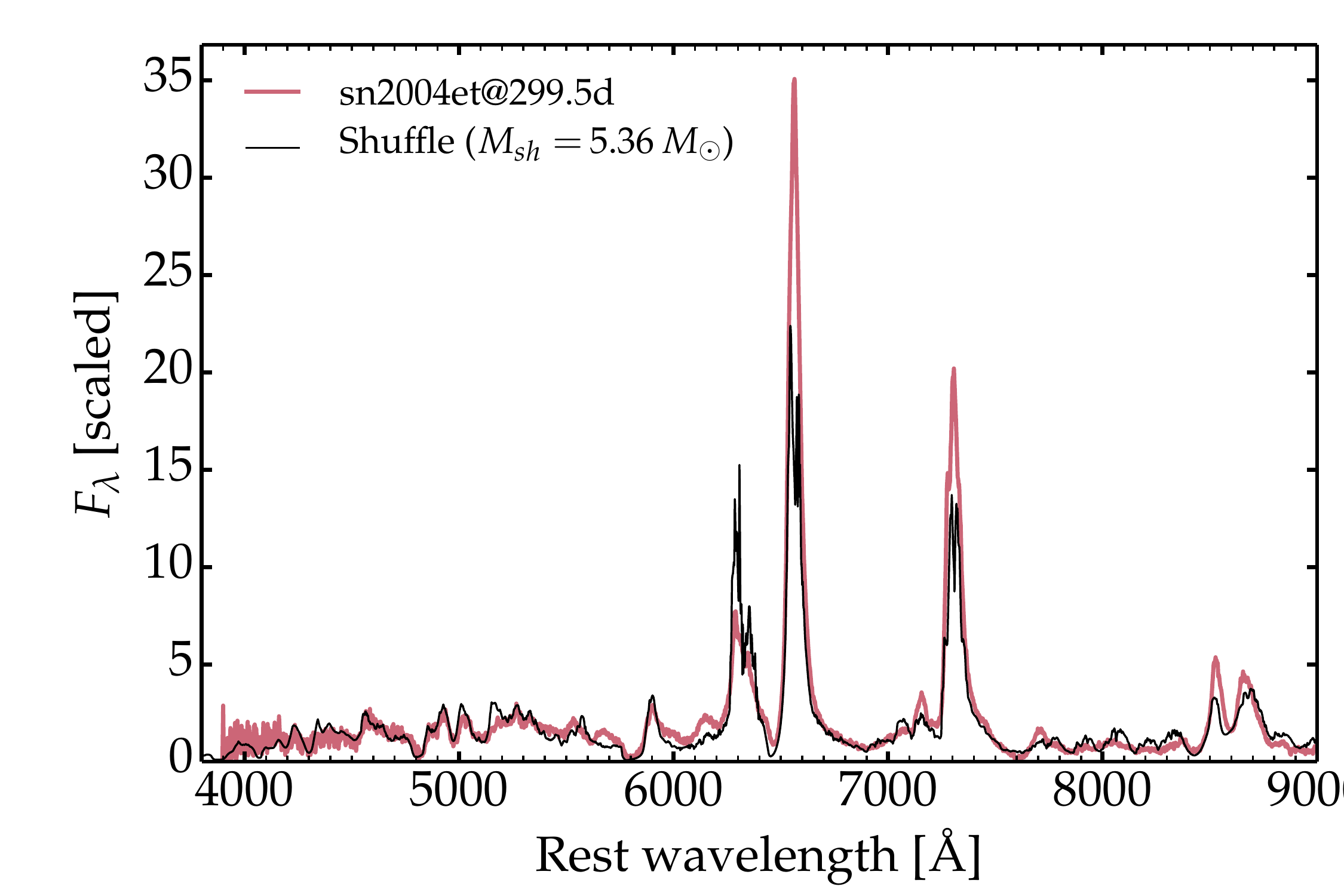}
\vspace{-0.6cm}
\caption{Comparison between the observations (corrected for reddening and redshift) of SN\,2004et at 299.5\,d after explosion with the shuffled-shell model with $M_{\rm sh}=5.36$\,\msun\ at 345\,d. We used a reddening of $E(B-V)=0.3$\,mag, and took the distance of 7.73\,Mpc from \citet{vandyk_17eaw_19}. With these parameters, the model is within 0.1\,mag of the $V$-band photometry of SN\,2004et at 350\,d, as reported by \citet{sahu_04et_06}. Here, model and observation are normalized at 6800\,\AA.
\label{fig_04et}
}
\end{figure}

Figure~\ref{fig_04et} shows a comparison between our shuffled-shell s15A model with $M_{\rm sh}=5.36$\,\msun\ and the observations of SN\,2004et. Model and observation are only offset by 0.1\,mag in $V$-band magnitude after correction for distance and reddening. The agreement in the spectral properties is also satisfactory, suggesting that model s15A is a suitable representation of the ejecta properties of SN\,2004et. One discrepancy is the lack of H$\alpha$ emission at low velocity. When we reverse the shell stacking order, more H is present in the inner ejecta but this causes only a modest increase in the H$\alpha$ strength (see Fig.~\ref{fig_spec}). Allowance for clumping might help resolve this small discrepancy.

\section{Discussion of pros and cons of the method}

The complex 3D structure of core-collapse SN ejecta, which typically exhibit asymmetry from small to large scales, would require high spatial and angular resolution together with 3D nonLTE radiative transfer. This is beyond current capabilities, which warrants the development of simplified approaches. Our approach has several caveats, but these are offset by the benefits.

At intermediate nebular times of a few hundred days, $\gamma$-rays have a large mean free path, and hence the arrangement of \nifs-rich material in shells, blobs, or fingers is not crucial for determining the energy deposition. With time passing,  $\gamma$-ray escape increases until the dominant  power source arises from positrons. In between these regimes, the adopted shell arrangement may influence the results, and the limitations of the approach may become more apparent. The transition from full $\gamma$-ray trapping to positron escape occurs on a time scale that varies considerably between SN types. For modeling SNe II at $\sim$\,300\,d or SNe Ibc at $\sim$\,150\,d, this is irrelevant so our technique should be robust in these cases. For the faster expanding SN Ia ejecta, $\gamma$-ray escape occurs much sooner. However, their chemical segregation is less severe than in core-collapse SNe, with the dominance of \nifs-rich material and Si-rich material (unburnt material plays little role at nebular times). Tests are needed to investigate the influence of the adopted shell arrangement for each SN type.

The radiation from one shell is modified as it crosses other shells. Blanketing below 5000\,\AA\ will lead to some frequency redistribution, in particular at earlier times. The arrangement in shells may overestimate this reprocessing since emission from inner shells is forced to cross all overlying shells before escape, while a more realistic configuration with fingers or blobs could facilitate this photon escape. But if SN ejecta are made of numerous blobs and fingers covering all sight lines, the difference with our simple radial shell arrangement might be small.

Our approach is fundamentally 1D and the emerging radiation is the same for any distant observer. This assumption will break down in the presence of large-scale asymmetries. Our technique, for example, will not capture the essential physics if the \nifs\ distribution is offset to one side. This could influence where the $\gamma$-ray power is absorbed, perhaps favoring some shells over other shells (for example, the Si-rich material might also be preferentially present on the same side as the \nifs, while the O-rich material could lie on the opposite side). We note that asymmetry will not affect the total flux in optically-thin lines (for the same decay power absorbed), but only their line profile.

Although presently ignored, we need to allow for compression (clumping) and depression (rarefaction) of the various shells in order to mimic the bubble effect caused by \nifs\ heating and the associated compression of the surrounding material. This will influence the ionization of the gas, for example enhancing the recombination in compressed regions (e.g., Ca in the H-rich material).

\section{Conclusions}

We have presented a new technique for the treatment of chemical mixing in SN ejecta for the purpose of radiative-transfer modeling. It relies on the shuffling in mass space (or equivalently velocity space) of the pristine ejecta shells produced in a 1D explosion model having reached homologous expansion. Each shell retains its original composition, thus no microscopic mixing  is introduced, while the shuffling of shells mimics the process of macroscopic mixing in velocity space. Because the method is 1D, it is amenable to detailed steady-state nonLTE radiative transfer calculations in grid-based codes. The need for several hundred grid points increases the memory requirements and the computing times by a factor of a few tens. To circumvent this hurdle, the calculation for a given explosion model is done in steps. First, we obtain a converged model using the standard mixing approach. The solution is then used for the initial guess on level populations, temperature, electron density etc for a shuffled-shell model using a Doppler line width set to 50\,\kms\ (for discussion on the role of the line Doppler width, see \citealt{DH20_neb}). Once converged, this model is then rerun with a line Doppler width fixed at 10\,\kms. The whole process takes several days, which is affordable for a steady-state calculation : only one model (rather than a full time-sequence of models) is needed to produce a spectrum at a given SN age.  The method requires some care but is straightforward to implement and can be applied to any SN type, either a core collapse or a thermonuclear explosion.

 With this more physically-consistent mixing technique, it becomes possible to perform reliable nebular-phase calculations for all SN types with \cmfgen. It is also of interest to test the impact of this different mixing technique during the photospheric phase, for example to assess the role of He\one\ line excitation in SNe Ib, or the behavior of H recombination during the plateau phase of SNe II-Plateau.

\begin{acknowledgements}

This work was supported by the "Programme National de Physique Stellaire" of CNRS/INSU co-funded by CEA and CNES. Hillier thanks NASA for partial support through the astrophysical theory grant 80NSSC20K0524. This work was granted access to the HPC resources of  CINES under the allocation  2019 -- A0070410554 made by GENCI, France. This research has made use of NASA's Astrophysics Data System Bibliographic Services.

\end{acknowledgements}

\newpage

\appendix

\section{Preparation of the initial conditions for the \cmfgen\ calculation with the new mixing technique}

To prepare the initial model for \cmfgen, we used the {\it unmixed} s15A ejecta model at 345\,d computed with \kepler\ \citep{WH07}, whose undecayed composition is shown in the left panel of Fig.~\ref{fig_prog_comp}. So, unlike numerous studies in which we used a combination of our own stellar evolution and explosion models (as in, for example, \citealt{dessart_snibc_20}), the initial conditions for the present \cmfgen\ calculations are based exclusively on that \kepler\ model s15A at 345\,d. Obviously, the new mixing technique can be applied to any unmixed coasting ejecta model, whatever its origin, and could also be used for calculations at other times during the coasting phase (i.e., when dynamical effects have abated).

At such a late time, the ejecta is essentially in homologous expansion (the current radius $R$ of each mass shell is huge relative to its initial radius), that is the velocity $V$ of each mass shell is simply  $V = R/ t$, where t is the time since explosion. For the \cmfgen\ input ejecta model, we enforced homologous expansion exactly, as usual  (this requirement arises from the numerical method; see details in \citealt{HD12}).

\begin{figure}
\centering
\includegraphics[width=\hsize]{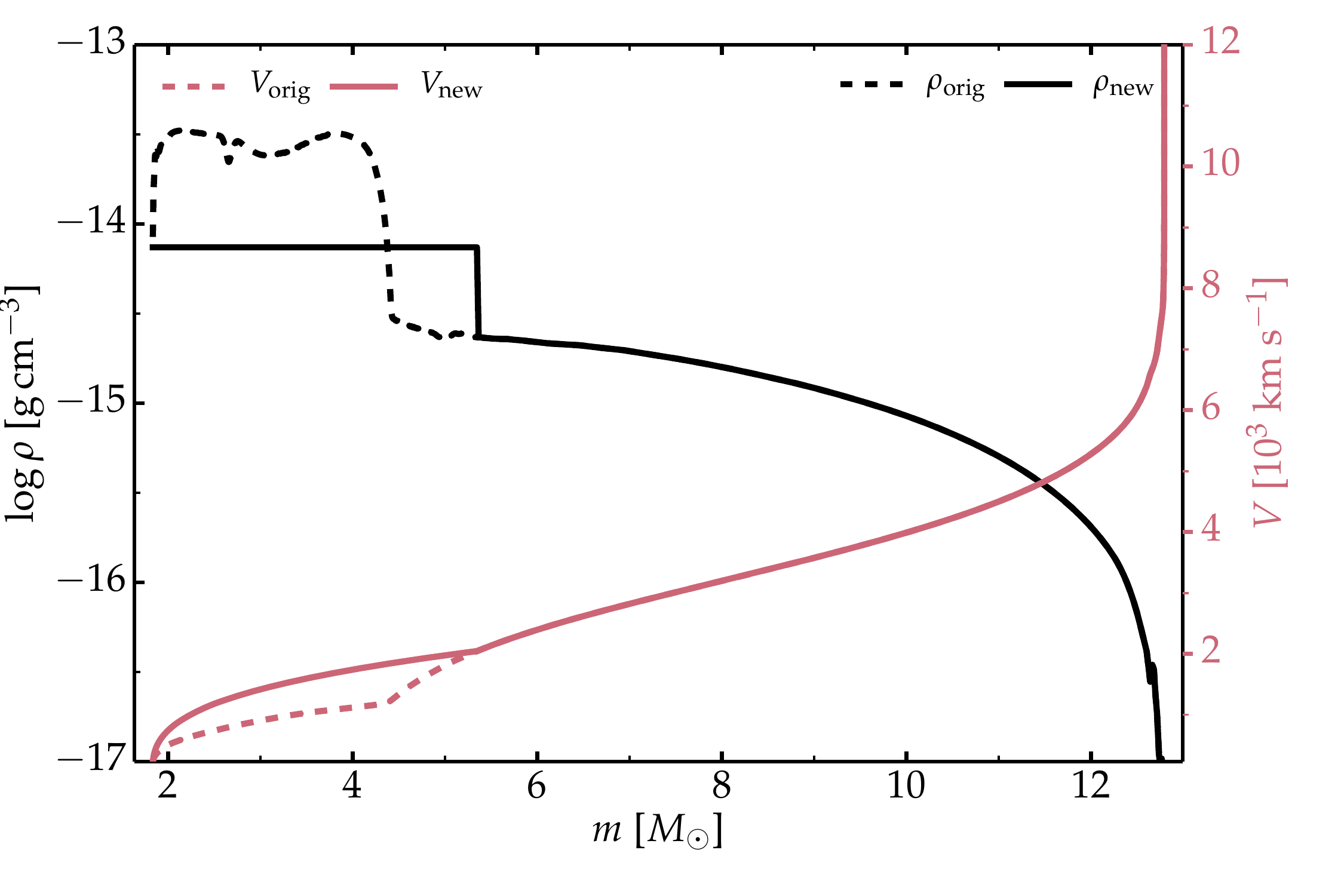}
\vspace{-0.6cm}
\caption{Density (black) and velocity (red) profiles versus Lagrangian mass are shown for the original ejecta model (dashed) and after the density has been reset in the inner ejecta regions bounded by $M_{\rm sh}$ (here chosen to be 5.36\,\msun).
\label{fig_m_v_rho}
}
\end{figure}

Before applying our mixing technique to the ejecta, we also suppress the density jump caused by the reverse shock formed at the He-core edge by forcing the density to be constant throughout the region we intend to mix (i.e., this density is just the total mass divided by the total volume of the ejecta between the innermost ejecta layer and $M_{\rm sh}$). Changing the inner density profile of 1D explosion models is routinely done by other radiative transfer modelers  (see, for example, Fig.~1 of \citealt{jerkstrand_04et_12}, where the density is similarly forced to be constant throughout the inner ejecta). This density jump is an artifact of the imposed spherical symmetry. Indeed, the Rayleigh-Taylor instability that grows at the He core edge after shock passage completely erases this jump (see, for example, Fig.~8 of  \citealt{utrobin_99em_3d_17} for a comparison between 1D and 3D predictions). Obviously, after changing the density in the inner ejecta regions, we have to recompute the Lagrangian mass, which is trivial. This change in density structure at low velocity affects the velocity versus mass, but only modestly since the inner ejecta contains a small fraction of the total kinetic energy. Indeed, the total ejecta kinetic energy increases by 3\%, while the kinetic energy within $M_{\rm sh}$, which is originally only 4\% of the total, is raised by 80\%. The velocity and density before and after this process are shown, versus Lagrangian mass, in Fig.~\ref{fig_m_v_rho}. This change in density profile (i.e., both versus radius and Lagrangian mass), is independent of the mixing step that we describe next.

 We then selected the region of the ejecta that we wished to mix. In the example shown in the right panel of Fig~\ref{fig_prog_comp}, we selected the region between the innermost layers at a Lagrangian mass of 1.83\,\msun\ (i.e. the remnant mass) and the Lagrangian mass $M_{\rm sh}$, here chosen to be 5.36\,\msun. The heart of the method is to swap a chunk $\delta M$ at some depth $M_1$ with the same $\delta M$ at some other depth $M_2$, both within $M_{\rm sh}$. Doing this obviously conserves mass (and the yields of all isotopes) since we are merely swapping chunks of identical mass between two different locations. It requires no change to the density profile (one does not even need to know the density structure to do this operation). The velocity structure is obviously unchanged in the process since it is uniquely defined by homology. Because the velocity increases with Lagrangian mass, this shuffling of shells leads to a shuffling in velocity space, which is precisely what we aim to achieve so that, for example, \nifs\ becomes present at higher velocities \citep{wongwathanarat_15_3d}.

There are many ways to design the shell swapping. One could, for example, choose to start at the ejecta base and swap a given $\delta M$ with the same $\delta M$ located at $M_{\rm sh}$, and then proceed until we reach $M_{\rm sh}/2$ (i.e., halfway through the regions to mix). But this is not desirable because some shells are thin and others thick. So, for example, with an adopted $\Delta M$ of a few 0.1\,\msun, all the \nifs\ would end up just below $M_{\rm sh}$. This situation would capture very poorly the predictions for \nifs\ mixing, which suggest the ubiquitous presence of \nifs\ from low to high velocities \citep{wongwathanarat_15_3d}. Similarly, these simulations suggest that blobs and fingers of distinct composition  (i.e., not just the \nifs\ material from the \nifs-rich shell) are thoroughly mixed within an extended region of the ejecta.

Hence, our choice of shell shuffling was driven by the need to shuffle {\it all} mass shells, so that material from the thin \nifs-rich shell and the thicker O-rich shell get mixed by comparable amounts. This way, fractions of the corresponding shells are made present in multiple regions of the ejecta.  Our approach was therefore to split each of the eight dominant shells (concisely tagged H/He, He/N, He/C, O/C, O/Ne/Mg, O/Si, Si/Ca, and Fe/He) into three equal parts and shuffle these 24 sub-shells within the same region limited by $M_{\rm sh}$. Starting from the ejecta base, we cycle three times through these eight shells,  placing each sub-shell on top of the previous one. When we are done, we have reached the Lagrangian mass $M_{\rm sh}$, beyond which the ejecta is unchanged from the original \kepler\ model. The result is similar to what is shown in the left panel of Fig.~\ref{fig_prog_comp} but now the pattern of eight shells is repeated three times with all eight shells being one third of their original mass but appearing at three different locations (this is most easily seen for the massive O-rich shell, colored in olive in Fig.~\ref{fig_prog_comp}). This shuffling (which is equivalent to swapping chunks of mass between two different locations) thus preserves the ejecta mass and kinetic energy but introduces the macroscopic mixing we need. There is no microscopic mixing between shells.

 To facilitate the procedure and the \cmfgen\ modeling, each of the eight shells are made homogeneous initially to avoid having additional composition gradients within shells. A very small mixing is also applied to all species to reduce the composition gradient at each shell interface in the original 1D model (see left panel of Fig.~\ref{fig_prog_comp}).

\end{document}